\documentclass[sigconf]{acmart}

\AtBeginDocument{%
  \providecommand\BibTeX{{%
    Bib\TeX}}}

\usepackage{tcolorbox}
\usepackage{textcomp}
\usepackage{caption} 
\usepackage{booktabs}
\usepackage{tabularx}
\usepackage{xcolor}
\usepackage{url}
\usepackage{booktabs}
\usepackage{enumitem}
\usepackage{tcolorbox}
\usepackage{colortbl} 
\usepackage{hyperref}
\usepackage{cleveref}
\usepackage{listings} 
\usepackage{graphicx}
\usepackage{subcaption}
\usepackage{framed}
\def\BibTeX{{\rm B\kern-.05em{\sc i\kern-.025em b}\kern-.08em
    T\kern-.1667em\lower.7ex\hbox{E}\kern-.125emX}}
\definecolor{ghostwhite}{rgb}{0.97, 0.97, 1.0} 

\crefformat{section}{\S#2#1#3}
\Crefformat{section}{\S#2#1#3} 

\newcommand{\code}[1]{\texttt{#1}}

\setcopyright{none}
\renewcommand\footnotetextcopyrightpermission[1]{}

\usepackage{listings}
\usepackage{multirow}
\usepackage{xspace}
\newcommand*{\eg}{e.g.,\@\xspace} 
\newcommand*{\ie}{i.e.,\@\xspace} 

\usepackage{float} 
\begin{document}

\title{Hot Fixing in the Wild}

\author{Carol Hanna}
\email{carol.hanna.21@ucl.ac.uk}
\orcid{0009-0009-7386-1622}
\affiliation{%
  \institution{University College London, UK}
  \country{} 
}

\author{Karine Even-Mendoza}
\email{karine.even_mendoza@kcl.ac.uk}
\orcid{0000-0002-3099-1189}
\affiliation{%
  \institution{King's College London, UK}
  \country{} 
}

\author{W.B. Langdon}
\email{w.langdon@cs.ucl.ac.uk}
\orcid{0000-0002-6388-4160}
\affiliation{%
  \institution{University College London, UK}
  \country{} 
}

\author{Mar Zamorano López}
\email{mar.zamorano@pdi.atlanticomedio.es}
\orcid{0000-0002-8872-4876}
\affiliation{%
  \institution{\mbox{Universidad del Atlantico Medio, Spain}}
  \country{} 
}

\author{Justyna Petke}
\email{j.petke@ucl.ac.uk}
\orcid{0000-0002-7833-6044}
\affiliation{%
  \institution{University College London, UK}
  \country{} 
}

\author{Federica Sarro}
\email{f.sarro@ucl.ac.uk}
\orcid{0000-0002-9146-442X}
\affiliation{%
  \institution{University College London, UK}
  \country{} 
}
\renewcommand{\shortauthors}{C. Hanna et al.}
\newcommand{\ch}[1]{\textcolor{orange}{#1}}

\begin{abstract}
Despite the operational importance of hot fixes, large-scale evidence on how they reshape routine maintenance workflows, particularly in the era of autonomous coding agents, remains limited. 
We analyse hot fixes present in over 61\,000 GitHub repositories from the 
Hao-Li/AIDev dataset and find 
consistent patterns of urgency:
reduced collaboration (typically a single contributor), smaller and more targeted changes (median 2–3 commits and files, with <10 line modifications), limited review (often fewer than two reviewers), and substantially fewer test file modifications than regular bug fixes, consistent with their urgency-driven character.
Leveraging the same urgency contexts, we examine differences between human- and AI-agent-authored hot fixes, revealing over 10 distinct repair behaviours,
thus offering insights into future human–automation collaboration for hot fixing.
Our study is the first to empirically analyse hot fix code changes at scale using a repository-level operationalisation of urgency.
The comparison of human and agent behaviours delineates their distinct characteristics, providing a foundation for understanding hot fixing in real-world practice.

\end{abstract}



\maketitle
\pagestyle{plain}

\section{Introduction}
Software systems inevitably encounter failures in production, yet not all failures carry the same 
operational weight \cite{hanna2024lirev,10.1145/3542929.3563482,10.1093/cybsec/tyab023,HPC:NotHotFix:10174003,10.1145/3338906.3338916,DoSinOS:5010241}.
Some defects cause outages \cite{10.1145/3338906.3338916,DoSinOS:5010241}, security exposures \cite{10.1093/cybsec/tyab023}, or severe user-facing disruptions that require immediate attention \cite{10.1145/3542929.3563482}. 

In practice, these high-stakes, high-urgency situations often take the form of \textit{hot fixes} as defined by Hanna et al. \cite{hanna2024lirev}: unplanned improvements to specific time-critical issues deployed to software systems in production.
Typically, these fixes are developed under time pressure and focused on mitigating symptoms rather than delivering a fully engineered permanent solution, and may bypass or ignore elements of the standard software development lifecycle.
Since correctness and completeness are temporarily de-prioritised in favour of rapid recovery, hot fix behaviour differs substantially from routine bug-fixing practice~\cite{hanna2025behind}. 

Patch classification has been widely studied through taxonomies, scoring, and prioritisation strategies~\cite{10.1007/s10207-025-01164-3,9970537,10.1145/3524459.3527343}, 
thus framing urgency primarily as a detection, classification or ranking problem for patch management rather than characterising how repair behaviours and workflow dynamics change under urgent conditions. 
Capturing hot fix scenarios,
in open repositories remains challenging. 
Issue trackers' labels, such as priority and severity, are rarely consistent across projects, as they may vary widely in their terminology and triage processes~\cite{10298540}.
Even with accurate labels, they alone do not imply such crisis-driven intervention; for example, HPC performance bugs (\eg{} inefficient implementations) may be severe yet typically do not fall within the scope of hot fixes \cite{HPC:NotHotFix:10174003}. 
Consequently, determining whether an issue constitutes a true hot fix scenario requires looking beyond explicit severity annotations to more contextual and behavioural cues.
Whether an issue reflects a time-critical production failure is often encoded implicitly in the language of the report, the urgency expressed by users, and the temporal dynamics of its resolution.
This ambiguity poses a challenge for studying hot fixes systematically at scale~\cite{hanna2025hotbugsjarbenchmarkhotfixes}.
We demonstrate this through a case study that extracts hot fixes from projects. We then leverage it for our main analysis, characterising repair behaviour under urgency for human developers and autonomous coding agents.

Autonomous coding agents are becoming active participants in software development teams.
The AIDev dataset~\cite{li2025aidev} captures over 61\,000 repositories and involves more than 47\,000 developers, where five popular autonomous coding agents propose, modify, and submit \textit{pull requests} (PRs) alongside human developers, capturing real-world patching activity in the wild.
The AIDev dataset spans heterogeneous agent workflows, ranging from human-in-the-loop, locally invoked coauthoring systems such as Claude Code to more autonomous, GitHub-integrated agents such as GitHub Copilot and Devin.
While recent analyses have examined agent productivity~\cite{10.1145/3633453}, review dynamics~\cite{watanabe2025useagenticcodingempirical}, and general patch quality~\cite{10992485}, little is known about how these agents behave in time-critical contexts.
Hot fixes represent a demanding subset of software maintenance: they require fast responsiveness to breakages that may occur without complete information.
Whether autonomous agents demonstrate such capabilities is an open, yet important question,
especially when hot fixing. 
Thus, we pose the following research question:
\textbf{\textit{How do autonomous agents and humans behave when addressing time-critical issues, compared to non-critical ones?}}

We examine two complementary axes: (i) comparing developer behaviour during hot fixes with behaviour during routine bug fix maintenance, and (ii) contrasting human-driven hot fixes with autonomous coding agent-authored hot fixes.
%
We find that hot fixes exhibit distinctive structural patterns, including fewer collaborators (typically one), smaller and more targeted code changes (less than ten line modifications), and reduced test interactions; and autonomous agents behave differently from humans, tending to generate narrower, more atomic fixes while humans apply broader changes with more deletions and iterative revisions.
Our paper makes the following contributions:
\begin{itemize}
    \item An empirical study of hot fix characteristics in large-scale code;
    \item First investigation into autonomous coding agent behaviour in the context of hot fixing;
    \item Discussion of future avenues of research on hot fixing based on our empirical results.
\end{itemize}


\parskip=1pt
\noindent{\textbf{Paper Structure.} }
We present our case study on automated hot fix classification in~\cref{casestudy}, the experimental design for answering our research question in~\cref{methodology}, results in~\cref{results}, and discussion in~\cref{discussion}.
The related work is described in~\cref{relatedwork}, whilst the limitations and future work are given in~\cref{limitationsAndFutureWork}, before we conclude in~\cref{conclusions}.

\parskip=1pt
\noindent{\textbf{Data Availability.} } 
We make all scripts for reproducing our results publicly available at:
\textcolor{blue}{\href{https://doi.org/10.5281/zenodo.17953526}{10.5281/zenodo.17953526}}.

\section{Case Study: Automated Hot Fix Classification}
\label{casestudy}
\noindent{\textbf{Motivation.} }
Despite the operational importance of hot fixes at scale, empirical evidence on how such contexts reshape maintenance behaviour remains limited. Without such evidence, claims about urgency-driven development, an activity of paramount importance for software reliability, remain largely anecdotal or confined to small-scale industrial case studies~\cite{hanna2025behind}.

This gap becomes more consequential with the emergence of autonomous coding agents. Characterising hot fixes at scale is therefore a necessary step of the foundation for analysing human-agent interaction under urgency. We begin with a feasibility case study in which we use a lightweight \textit{Large Language Models} (LLM) assisted triage to support our main analysis (\cref{methodology}). This process is used to identify candidate hot fixes, which are then manually validated. The resulting hot fixes form the basis for answering the research question of this study.
\parskip=1pt
\noindent{\textbf{Pragmatic Solution.} }
To enable our empirical evaluation, we first identify hot fixes within the Hao-Li AIDev dataset \cite{li2025aidev}. We conduct an exploratory case study to assess whether LLM-based classification, combined with temporal heuristics, can reliably distinguish hot fixes from routine \textit{pull requests} (PRs). This validation step establishes the foundation for our subsequent large-scale behavioural analysis.

\label{method:rq1}
Since issue trackers rarely provide consistent severity or priority labels, and none of the repositories in our sample included standardised criticality indicators, we first perform LLM-based classification over AIDev's \mbox{\textit{issue.body}} as follows;
\begin{enumerate} [nosep,noitemsep,leftmargin=15pt,itemsep=0pt, topsep=0pt]
\item We implement the classifier using one-shot prompting on three LLMs to capture linguistic urgency cues; see~\autoref{tbl:ollama:versions}.

\item We then apply temporal filters to more accurately capture true hot fixes. 

\item We conduct a manual validation on a sampled subset to evaluate our approach accuracy in identifying time-critical issues. 
\end{enumerate}
Next, we detail the three steps.

\noindent\textbf{(1) Identifying Hot Fixes via LLM.}
We selected three small 
\linebreak[4]
\mbox{($\le 2.5$ Gigabytes),} locally deployable LLMs: Llama 3.2, Qwen, and Phi-4 via Ollama, run on a standalone computer\footnote{Intel® Core™ Ultra 7 155U×14 processor, 64GiB memory \& Ubuntu 22.04.5 LTS 64-bit.} with no GPU (\autoref{tbl:ollama:versions}).
Our choice is guided by two considerations. \textit{First}, 
the process must be applicable in resource-constrained, privacy-sensitive environments, and maintain reproducibility and long-term availability to mimic real developers' workflows,
where sending proprietary issues and commit text to external APIs may be impossible.
Thus, we select LLMs that run efficiently on developer workstations.
\textit{Second}, we select LLMs based on existing work in the domain of hot fixes (Llama~3~\cite{ssbse:hotcat:2025}), and to represent complementary design choices. 
Llama~3 and Qwen have demonstrated strong performance on general coding and software-engineering tasks, 
while Phi-4 is compact, relatively new, comes from a different LLM family (Microsoft vs.~Meta) and is suitable for classification. 
We considered other LLMs, \eg{} \code{qwen3:8b}, but they were too slow due to higher token requirements.
We leave to future work hardware configurations examination (CPU-only, single/multi-GPU), known to impact coding-related tasks' performance and accuracy in the wild~\cite{ASE:refuzzer:2025}. 

For each issue, we prompt the LLM to determine whether the description reflects a time-critical, production-impacting failure.
We implement prompting via a simple chat-completion call with a \textit{system message} (\textit{You are a software triage assistant. Only answer with: "critical" or "not\_critical"}), a single \textit{user message} as shown in Listing~\ref{prompt}, zero temperature, and a strict maximum of five output tokens to enforce a binary response. 
The prompt was refined iteratively and experimentally. We began with the formatting instruction, hot fix definition, and issue body, then progressively added specific guidelines and examples after inspecting early outputs.

\begin{table}[t!]
\centering
\small
\caption{Ollama LLM Details}
\begin{tabular}{@{}lcrr@{}}
\hline
\multicolumn{1}{c}{\textbf{Model}} &
\multicolumn{1}{c}{\textbf{Image ID}} &
\multicolumn{1}{c}{\textbf{Size}} &
\multicolumn{1}{c}{\textbf{Release date}} \\
\hline
\texttt{qwen2.5:3b-instruct} & \texttt{357c53fb659c} & 1.9 GB & Dec 2025 \\
\texttt{phi4-mini:latest}    & \texttt{78fad5d182a7} & 2.5 GB & Dec 2025 \\
\texttt{llama3.2:latest}     & \texttt{a80c4f17acd5} & 2.0 GB & Nov 2025 \\
\hline
\end{tabular}
\label{tbl:ollama:versions}
\vspace{-0.5 cm}
\end{table}
\begin{lstlisting}[
  caption={LLM prompt used for issue criticality classification.},
  label={prompt},
  basicstyle=\ttfamily\footnotesize,
  breaklines=true,
  breakatwhitespace=true,
  columns=fullflexible, float
]
Determine if the following GitHub issue describes a time critical issue that should be considered a hot fix.
A hot fix is an unplanned improvement to a specific time-critical issue deployed to a software system in production.
Classify an issue as "critical" ONLY if it clearly describes a problem that affects production users NOW, AND requires immediate action outside normal release cycles.
Classify an issue as "not_critical" if ANY of the following apply:
- The issue does NOT mention production, real users, or a deployed environment.
- The severity is unclear or speculative.
- The problem can be scheduled normally.
- It is a minor bug, cosmetic, enhancement, or development-only failure.
- Impact is unknown, the issue is missing context, or your decision is borderline.
Issue body:    {body}
\end{lstlisting}

\noindent\textbf{(2) Time Stamp Filter. }
Stage (1) outputs either \textit{"critical"} or \textit{"non-critical"}.
To further refine the candidate hot fix issues set, we then incorporated temporal metadata. Specifically, we applied two temporal filters based on heuristics from previous work~\cite{hanna2025hotbugsjarbenchmarkhotfixes}.
First, examining the time delta between each PR’s \textit{created\_at} timestamp and 
its corresponding issue's \textit{created\_at} timestamp to identify PRs that were opened urgently after the issue's identification.
The second filter checks the differences between the PR's \textit{created\_at} and \textit{closed\_at} timestamps,
to identify PRs that were opened and resolved quickly, reflecting the rapid turnaround characteristic of real hot fixes. After experimenting with several time delta thresholds, we found that cutoffs of 12 and 24 hours respectively
provided the best balance between minimizing false positives (routine issues closed quickly) and false negatives (genuinely urgent issues with slightly delayed closure).
We then link issues to the PRs intended to address them, using the \textit{related\_issue} table that maps \textit{issue\_id} to \textit{ pr\_id}. This allows us to reconstruct the repair chain for each issue.
%
We conduct data sanitization on the resulting data,
removing problematic characters (\eg{}, ",\textbackslash r\textbackslash n) inside the title and description fields of each PR and issue.
All other content is preserved unchanged, so the textual information remains as is.
This step focuses only on structural issues, leaving more sophisticated sanitisation for future work.\looseness=-1


\noindent\textbf{(3) Manual Validation. } 
\label{manualeval} To ensure reliability, for each of the three LLMs results, four of the authors (each with at least 5 years of industrial software development experience) independently manually inspected a 20\% random sample from the issues identified as hot fixes by the LLMs to evaluate if each was actually a hot fix. Then, in a meeting, notes were compared to resolve cases where they had disagreed. These disagreement resolution conversations led to subsequent discussions on the contextual factors not available to the LLM-based classifier (from GitHub or the website associated with the repository) affecting hot fix classifications.

\begin{table}[t!]
\centering
\small
\caption{Automated issue filtering and PR count summary}
\label{tab:issuePRcounts}
\begin{tabular}{@{}l@{ }llrrr@{}}
\toprule
& & \multicolumn{2}{r}{\textbf{Llama3.2}} & \textbf{phi4mini} & \textbf{qwen}\\
\hline
\multirow{2}{*}{\textbf{Issues}} 
&Non-Critical & LLM & 4347 & 4530 & 4588\\\cline{2-6}
& \multirow{2}{*}{Critical} &  LLM + no temporal& 1348 & 425 & 148\\
&  & LLM + temporal & 269 & 105 & 33\\
\midrule
\multirow{6}{*}{\textbf{PRs}} & \multirow{3}{*}{Non-Critical} & Bots & 397 & 425 & 433\\
& & Human & 4258 & 4392 & 4443\\\cline{3-6}
& & Total & 4655 & 4817 & 4876\\\cline{2-6}

& \multirow{3}{*}{Critical} & Bots & 41 & 13 & 6\\
& & Human & 229 & 99 & 31\\\cline{3-6}
& & Total & 270 & 112 & 37\\

\bottomrule
\end{tabular}
\end{table}
\begin{table}[tb]
\centering
\small
\caption{Examples of Ambiguous Issues Assessed}
\label{tab:hotfix_examples}
\begin{tabular}{@{}p{1.4cm} p{6.7cm}@{}}
\hline
\textbf{Issue} & \textbf{Key Observations}  \\ \hline
\texttt{microsoft/ react- native- windows- samples \#1054}&
\textbf{Not a hot fix. }Initially flagged as a potential hot fix due to a package dependency update and rapid patching. However, milestone and release inspection showed the change was scheduled for a future milestone rather than an immediate production release, indicating no urgent production impact.  \\ \hline
\texttt{GitHub Issue \#38880}&
\textbf{Not a hot fix. }Despite urgency-like language (\eg{} references to “many GitHub users”), the issue concerns a recreational game with no deployment context or live system dependency. As such, it does not affect production environments.  \\ \hline
\texttt{stellar/ rs-soroban- sdk \#1500}&
\textbf{A hot fix. }Appears to be a small and quickly resolved without urgent labels. Closer inspection reveals it affects Soroban smart contract compilation, causing failures in nightly builds due to a value conversion error. In Stellar’s deployment context, this can block contract deployment and introduce security risks.\looseness=-1
\\ \hline
\end{tabular}
\end{table}
\parskip=2pt
\noindent\textbf{Case Study Findings. } 
\autoref{tab:issuePRcounts} summarises the results of using three different LLMs to classify non-critical and critical issues. We observe that using the time between issue/PR creation and closure or merge as an additional filtering criterion (\ie{} row ``LLM + temporal'' in \autoref{tab:issuePRcounts}) substantially reduces the number of the issues classified as critical, indicating gaps between urgent wording and action in practice. 
Overall, \code{llama3.2} marked more issues/PR as critical, followed by \code{phi4-mini} and \code{Qwen2.5}.

The manual evaluation required over three hours of joint analysis and discussions to resolve disagreements. We first attempted resolution using only issue and PR descriptions; when this was insufficient, external context was incorporated (\autoref{tab:hotfix_examples}). In total, all 40 disagreements in the 20\% sample were resolved, out of which 12 required external resources. After resolution, overall agreement between human reviewers and LLMs was 0.37. Agreement varied across LLMs: \code{qwen2.5} (most conservative) fully aligned with reviewers, \code{phi-4-mini} (intermediate) agreed in $\approx 50\%$ of cases, and \code{llama3.2} (most flexible) had a similar disagreement rate (49\%) but all 12 externally resolved cases originated from \code{llama3.2}’s predictions.\looseness=-1

Our discussion highlighted a core limitation of content-only hot fix identification (\ie{} using only the PR and issue text for classification): even when an issue/PR sounds urgent, it may be found not to be a hot fix once we inspect the surrounding context. 
We have identified during our discussions the following external parameters that affected our decision not to classify these PR as a hot fix:
(1)~\textit{Project README.md} to indicate whether the system is intended to be production-grade with users; 
(2)~\textit{project's deployment} including either download page of its binaries or active hosted site to indicate if this is a live project; (3)~\textit{merge information} to check whether the PR fix was merged into main/production, waiting for future release or milestone, or a draft awaiting merge
(4)~\textit{repository maturity} indications such as commit and contributor count, release history, and age of repository; and 
(5)~\textit{labels tags} in PR and issue (enhancement or request vs urgent or blocking). Examples are in \autoref{tab:hotfix_examples}.
\looseness=-1

%

\section{Analysis Approach for Hot Fix Behaviours}
\label{methodology}

Our goal is to analyse how humans and autonomous coding agents behave when addressing time-critical production issues. We focus on a subset of issues resembling hot fix scenarios, as defined in \cref{casestudy}, within the Hao-Li AIDev ecosystem~\cite{li2025aidev}. 
The hot fixes observed in the dataset are written most commonly in TypeScript, Python, C\#, Go, and Rust.
This corresponds with the general distribution of programming languages in the dataset under study.

We structure our analysis along two dimensions: (i) linking issues to their \textit{pull requests} (PRs), and (ii) distinguishing between agent-generated and human-authored PRs. The result of the analysis comprise a \textit{quantitative comparison} of hot fix and routine bug fix PRs using PR level characteristics (contributors, commits, reviewers, files/LOC changed, test-file contributions, merge status), and a \textit{qualitative linguistic analysis} of PR text across authorship (human vs agent) of hot fixes.
%
%

\noindent\textbf{Analysis. }
\label{method:hotfix:char}
We link issues classified as time-critical to their associated PRs using 
AIDev's \textit{related\_issue} table. We then extract PR-level characteristics by joining several tables: \textit{pr\_commits} and \textit{pr\_commit\_details} for commit counts, files touched, and line additions/deletions; \textit{pr\_reviews} for reviewer counts; and the repository for identifying dominant languages. 
Further, we detect test files modifications by scanning commit filenames, and apply the same procedure to non-critical issues, and 
to human- and agent-authored PRs, enabling direct comparison across hot fix and regular maintenance activities.
We compare linguistic characteristics of human and bot authored hot fix PRs, by applying a bag-of-words approach with word-cloud visualisation. 
We aggregate PR titles and descriptions into two corpora: human-authored and bot-authored. The produced visualisation contains the 40 most frequent terms in each corpus using \code{WordCloud} and \code{Matplotlib}. 


\section{Results: Hot Fix Characteristics}
\label{results}

\noindent\textbf{Quantitative Results. } 
\autoref{tab:hotfixVSbugfix} (col. \textit{PR Fix Type}) present the results of 
comparing hot fixing characteristics with regular bug fixing. 

We find that \textbf{(1)} the \textit{number of contributors} involved in a hot fix is consistently lower than for regular bug fixes with a maximum of 5 contributors for hot fixes and 13 for regular fixes across the evaluated PRs. This result aligns with a previous human study examining industrial hot fixing practices~\cite{hanna2025behind}, where such activities were typically described as involving a small group, often consisting of a team lead and an on-call developer.
\textbf{(2)} This pattern is also reflected in the \textit{number of code reviewers} assigned to hot fix \textit{pull requests} (PRs), which on average is lower than for regular bug fixes. The median number of reviewers for regular fixes is 2.3 and for hot fixes is 2. However, some regular fix PRs had up to 16 reviewers whereas the maximum we found for hot fixes was 5. This suggests an expedited review process designed to ensure that the patch can be deployed as quickly as possible. 
\textbf{(3)} Hot fixes tend to contain fewer \textit{commits}, involve significantly fewer changed files, and include far fewer lines added or removed
For hot fix pull requests found with the qwen2.5-3b: a mean of 2.7 commits, 3.9 changed files, 25.7 lines added, and 9.3 lines removed in comparison with mean values of 4.9, 27.7, 90, and 54.4 for regular pull requests respectively.
This aligns with the expectation that a hot fix targets a specific issue in production, rather than performing large-scale refactoring. Hot fixes are inherently focused interventions aimed to resolve a singular critical problem. 
\textbf{(4)} Hot fix PRs are less likely to include changes to \textit{test files} than standard bug fix PRs (a reduction of up to 24.68\%). This may be due to the time pressure under which hot fixes are developed, leading teams to bypass certain software testing best practices to enable rapid deployment. Finally, \textbf{(5)} hot fix PRs are \textit{merged} more often than regular bug fixes (up to 25\% more likely). 

Next, we separate hot fix PRs authored by human developers from those generated by autonomous coding agents, as shown in \autoref{tab:hotfixVSbugfix} (col. \textit{PR Contributor Type}).
The results of this analysis
reveal distinct patterns in collaboration, effort, and code modification behaviour. 
At present, \textbf{(6)} most hot fix PRs are human-authored (229 hot fix PRs vs 41 regular fix PRs).  
Human PRs tend to involve slightly less participants on average (mean of 1.2 vs 1.5 contributors for qwen2.5-3b), suggesting that bot-authored changes more frequently require additional discussion or coordination. Humans, by contrast, typically operate in a more self-contained manner, with their PRs involving fewer contributors. 
\textbf{(7)} \textit{Commit activity} also differentiates human and bot behaviour.
Human PRs generally contain slightly more commits (up to 0.9 more commits on average), reflecting a more iterative development process that includes revision and refinement before final submission. Bots typically generate atomic, single-intent changes, an indication of their procedural and predictable update patterns. 
\textbf{(8)} \textit{Review patterns} further highlight differences in how PRs are processed.
Bot PRs often involve only a single reviewer, suggesting the perception that bot-generated fixes are trivial. This reduced reviewer load may also indicate that bot PRs are narrower in scope and thus easier to evaluate. 
\textbf{(9)} The size of the \textit{code changes} reveals some of the most striking contrasts.
While bots tend to generate small and predictable modifications, they often introduce more lines of code on average than humans, 
particularly in the llama3.2 data, where on average bot code additions substantially exceed human ones (44 more lines per PR on average). 
However, humans are more likely to delete when hot fixing.
Human PRs also typically modify more files than bot PRs (4 more files on average for llama3.2), suggesting that human contributors address more complex or interconnected hot fixes. Bots focus on tightly scoped issues affecting few files, consistent with automated tooling that is optimized for changes narrow in scope.
\textbf{(10)} \textit{Test file interactions} present another interesting difference:
Bot PRs in both phi4-mini and llama3.2 touch test files at substantially higher rates than human PRs (up to 25\%), suggesting pipelines that proactively adjust tests alongside code changes. 
Finally, we observe that bots achieve relatively strong merge success rates, comparable to human ones.

\begin{figure}[tb]
    \centering
    \begin{subfigure}[b]{0.49\linewidth}
        \centering
        \caption{Human Pull Requests}
\includegraphics[width=\linewidth]{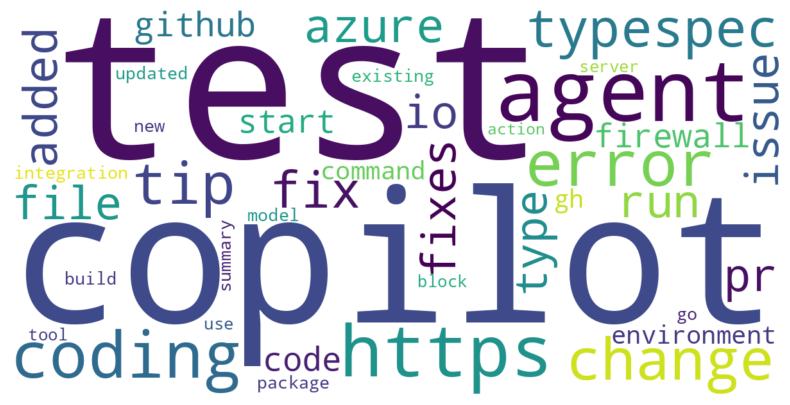}
        \Description{Two figures of word counts.}
        \label{fig:wordcloud:human}
    \end{subfigure}
    \hfill
    \begin{subfigure}[b]{0.49\linewidth}
        \centering
        \caption{AI Agent Pull Requests}
\includegraphics[width=\linewidth]{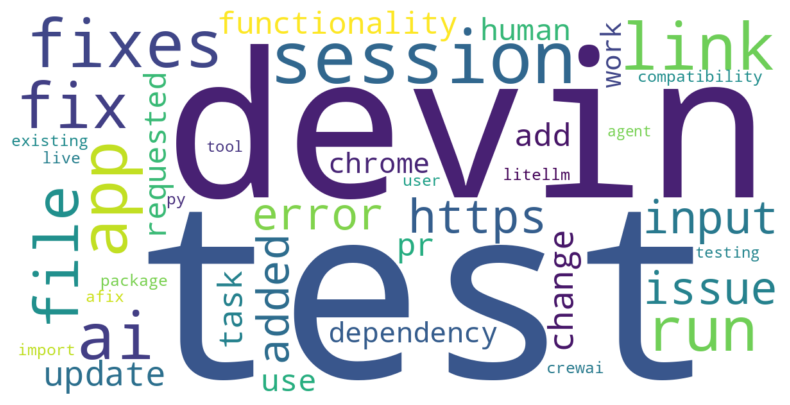}
        \Description{Two figures of word counts.}
 \label{fig:wordcloud:bot}
    \end{subfigure}
    \vspace{-0.2cm}
    \caption{Word clouds for human/bot-initiated Hot Fix PRs.}
    \Description{Two figures of word counts.}
    \vspace{0.1 cm}
\label{fig:wordcloud:comparison}
\end{figure}

\begin{table}[ht]

\centering
\small
\caption{Hot Fix (Bot and Human) Vs. Regular Bug Fix
\mbox{Characteristics} (Med, Mean; Min-Max).}
\label{tab:hotfixVSbugfix}
\scalebox{0.86}{
\begin{tabular}{@{}l@{}l@{}l@{ }l@{ }l@{ }l@{}}
\toprule
\textbf{Characteristic} &
\multicolumn{2}{l@{}}{\textbf{PR Fix Type}} &
\textbf{qwen2.5-3b} &
\textbf{llama3.2} &
\textbf{phi4mini} \\
&\multicolumn{2}{@{}r@{}}{\textbf{Contributor Type}}
\\
\midrule

\multirow{4}{*}{\textbf{\#Contributors}}
&  Regular &
& 1, 1.3 ; 1-13
& 1, 1.3 ; 1-13
& 1, 1.3 ; 1-13 \\

& Hot  & Total
& 1, 1.2 ; 1-2
& 1, 1.3 ; 1-5
& 1, 1.2 ; 1-2 \\

& & Humans
& 1, 1.2 ; 1-2
& 1, 1.2 ; \textbf{1-5}
& 1, 1.1 ; 1-2 \\

& & Bots
& \textbf{1.5}, \textbf{1.5} ; 1-2
& 1, 1.4 ; 1-3
& 1, 1.3 ; 1-2 \\
\midrule

\multirow{4}{*}{\textbf{\#Commits}}
&  Regular &
& \textbf{3}, \textbf{4.9} ; \textbf{1-30}
& \textbf{3}, \textbf{4.9} ; \textbf{1-30}
& \textbf{3}, \textbf{4.9} ; \textbf{1-30} \\

& Hot  & Total
& 2, 2.7 ; 1-10
& \textbf{3}, 3.9 ; \textbf{1-30}
& 2, 3.2 ; 1-15 \\

& & Humans
& 2, 2.9 ; 1-10
& \textbf{3}, 3.9 ; \textbf{1-30}
& 2, 3.3 ; 1-15 \\

& & Bots
& 2, 2.0 ; 1-3
& 2, 3.5 ; \textbf{1-30}
& 2, 2.6 ; 1-10 \\
\midrule

\multirow{4}{*}{\textbf{\#Reviewers}}
& Regular  &
& 2, \textbf{2.3 }; \textbf{1-16}
& 2, \textbf{2.3 }; \textbf{1-16}
& 2, \textbf{2.3} ; \textbf{1-16} \\

& Hot  & Total
& 2, 1.8 ; 1-3
& 2, 2.0 ; 1-5
& 2, 2.2 ; 1-4 \\

& & Humans
& 2, 1.9 ; 1-3
& 2, 2.1 ; 1-5
& 2, 2.2 ; 1-4 \\

& & Bots
& 1.5, 1.5 ; 1-2
& 1, 1.1 ; 1-2
& 1, 1.2 ; 1-2 \\

\midrule

\multirow{4}{*}{\textbf{Lines added}}
& Regular  &
& 4, 90; \textbf{0-237419}
& 4, 90; \textbf{0-237419}
& 4, 91; \textbf{0-237419} \\

& Hot  & Total
& 5.5, 25.7 ; 0-383
& 4, 69.2 ; 0-39116
& 2, 19.2 ; 0-1942 \\

& & Humans 
& 5, 24.4 ; 0-383
& 4, 66.0 ; 0-39116
& 2, 18.4 ; 0-1942 \\

& & Bots
& 6, 33.9 ; 0-378
& \textbf{17}, \textbf{110.2} ; 0-2571
& 6, 34.9 ; 0-378 \\

\midrule

\multirow{4}{*}{\textbf{Lines deleted}}
&  Regular &
& 1, 54.4; \textbf{0-237419}
& 1, 53.9; \textbf{0-237419}
& 1, 50.6; \textbf{0-237419} \\

& Hot  & Total
& 1, 9.3 ; 0-186
&\textbf{ }2, 74.1 ; 0-31937
& 1, 553 ; 0-90905 \\

& & Humans
& 1, 10.3 ; 0-186
& \textbf{2}, 72.8 ; 0-17314
& 1, \textbf{584} ; 0-90905 \\

& & Bots
& 1, 3.6 ; 0-61
& 1, 91.2 ; 0-31937
& 1, 4.7 ; 0-63 \\

\midrule

\multirow{4}{*}{\textbf{Files modified}}
&  Regular &
& \textbf{4}, 27.7 ; \textbf{0-1484}
& \textbf{4}, \textbf{28.3} ; \textbf{0-1484}
& \textbf{4}, 27.9 ; \textbf{0-1484} \\

& Hot  & Total
& 2, 3.9 ; 0-14
& 3, 16.4 ; 0-485
& 3, 9.6 ; 0-300 \\

& & Humans
& 2, 3.9 ; 0-14
& 3, 16.9 ; 0-485
& 2, 10.4 ; 0-300 \\

& & Bots
& 3, 4.2 ; 1-11
& 3, 12.9 ; 1-301
& 3, 4.2 ; 1-11 \\

\bottomrule

\multirow{4}{*}{\textbf{Includes tests}}
& Regular  &
& 54.42\%
& 53.68\%
& 53.57\% \\

& Hot  & Total
& 29.73\%
& 50.14\%
& 34.82\% \\

& & Humans & 29.03\%
& 47.50\%
& 31.31\% \\

& & Bots & 33.33\%
& \textbf{72.97\%}
& 61.54\% \\

\midrule
\multirow{4}{*}{\textbf{Merged}}
&Regular   & 
  & 45.30\% 
  & 43.68\% 
  & 44.84\% \\
  
& Hot & Total
  & 70.27\% 
  & 58.94\% 
  & 68.75\%  \\

&  & Humans 
  & \textbf{70.97\% }
  & 58.26\% 
  
  & 68.69\% \\
& & Bot
  & 66.67\% 
  & 64.86\% 
  & 69.23\% \\

\bottomrule
\end{tabular}
}
\end{table}

\label{sec:wordcloud:comparison}
\autoref{fig:wordcloud:comparison} shows word cloud visualisations of our use of English-language analysis in the PR bodies to identify systematic differences between human- and agent-authored hot fix PRs.
%
While both corpora heavily reference testing, their vocabularies differ in clear ways. 
\textit{Human-authored PRs} are dominated by terms like {\tt code} and {\tt Copilot} relating to assistant tooling,
nouns relevant to the PR such as
{\tt TypeSpec}, {\tt Azure}, 
and words around error framing differences, such as   
{\tt change}, {\tt updated} and {\tt added}.
In contrast, \textit{bot-authored PRs} use more workflow-related language such as 
{\tt session}, {\tt task} and {\tt link},
terms related to humans (\eg, {\tt user} and {\tt human}),
and PR aims, such as {\tt functionality} and
{\tt compatibility}. 
%
Together with the case study results, these results can inform feature-based machine learning predictors in future work on hot fixes and their characterisation.\looseness=-1


\begin{framed}
    \noindent\textbf{Takeaway:} When addressing critical issues, both humans and autonomous AI agents produce \textbf{smaller}, \textbf{faster}, and \textbf{more focused changes} than for non-critical tasks, with \textbf{fewer contributors},\textbf{ fewer files modified}, and\textbf{ expedited review processes}. Humans tend to handle hot fixes in a more self-contained and iterative manner, modifying more files and deleting more code, while bots generate narrowly scoped, atomic changes that often include test updates. Overall, critical issues amplify these differences: humans address complexity and interdependencies, whereas \textbf{autonomous agents concentrate on tightly scoped fixes rather than broader, interconnected changes.}
\end{framed}



\section{Discussion}
\label{discussion}

We provide the first large-scale empirical characterisation of hot fix practices in open-source software and the first direct comparison between human and autonomous AI agent behaviour in time-critical repair contexts.
Our findings show that hot fixes differ systematically from regular bug fixes in structure, collaboration patterns, and testing practices, and that these differences are further amplified when comparing human-authored and agent-authored \textit{pull requests} (PRs).
Across all three classifiers, hot fixes exhibit strong signatures of urgency, including fewer contributors, fewer commits, fewer modified files, fewer test files, and reduced review participation.
These results reinforce prior qualitative findings from industrial settings that hot fixes are typically handled by a small number of developers and prioritise rapid deployment over comprehensive engineering processes~\cite{hanna2025behind}.
The lower likelihood of modifying test files especially highlights the trade-off between speed and assurance.
This suggests that teams may consciously relax testing discipline to restore production stability as quickly as possible.
This behaviour reflects a pragmatic response to operational pressure, where mitigating immediate user impact outweighs long-term maintainability.\looseness=-1

We speculate that many hot fixes remain human-authored because they are often triggered by immediate alerts and handled under a stronger sense of personal responsibility, whereas delegating such issues to an agent would require formal issue creation and introduce additional response latency
When contrasting human and autonomous agent hot fixes, we observe distinct repair strategies.
Human developers tend to produce more iterative and broader patches, touching more files and deleting more lines of code.
This pattern suggests that humans engage more deeply with contextual dependencies and system-wide implications of failures, even under time pressure.
In contrast, autonomous agents generate more atomic and narrowly scoped changes.
This behaviour is consistent with their optimisation for isolated tasks and well-defined instructions.
These behavioural differences indicate that agents currently operate as precision tools rather than holistic responders to production incidents.
Interestingly, agents modify test files at a substantially higher rate than humans in hot fix scenarios.
This may reflect built-in pipelines or prompts that encourage test updates alongside code changes, even when human developers deprioritise such practices.
While this behaviour could improve short-term correctness, it also raises questions about whether agents fully understand the production urgency context or simply apply generic repair heuristics.
In urgent settings, test updates may be unnecessary or even risky if they delay deployment.
This highlights a potential misalignment between automated workflows and real-world operational priorities.

Despite these differences, agent-authored hot fixes achieve merge rates comparable to human-authored ones.
This suggests that autonomous agents can already participate meaningfully in urgent repair workflows, at least for narrowly scoped problems.
However, their reduced peer review attention may indicate either higher trust in triviality or lower scrutiny due to perceived low risk.
This remains unexplored in the scope of this study but raises an interesting point for investigation in the future given that hot fixes by definition operate in high-stakes environments where errors can have severe consequences.

Taken together, our results suggest that humans appear to manage complexity and interdependencies, whereas agents focus on localised corrections.
This division of labour points toward promising hybrid workflows in which agents assist with targeted fixes while humans retain responsibility for system-level reasoning and risk assessment.
Designing agents that can better recognise urgency, adapt testing and review strategies accordingly, and integrate contextual knowledge about deployment environments remains an important challenge for future research.
More broadly, our findings demonstrate that hot fixing is not merely faster bug fixing, but a qualitatively distinct mode of software maintenance with its own social and technical characteristics.
Understanding these patterns is essential as autonomous agents become increasingly embedded in production-facing development processes.

\section{Related Work}
\label{relatedwork}

Recent advancements in LLM-based agentic systems for software issue resolution have begun to pave the way toward automated software maintenance \cite{Jiang2025}. Despite these technical advances, the interaction between developers and such agents in real-world workflows remains under explored, particularly with respect to trust in the generated code's correctness and reliability \cite{roychoudhury2025agenticaisoftwarethoughts}.
Emerging studies have started to examine agent-generated PRs, including when they are merged or rejected \cite{alam2026aiagentinvolvedpull,pinna2026} and the reasons underlying their rejection \cite{ehsani2026aicodingagentsfail}. However, these investigations have largely focused on routine development scenarios and do not consider time-critical contexts such as emergency hot fixing.
Hot fixing is a distinct software engineering activity aimed to restabilise a software system after a time-critical failure \cite{hanna2024lirev}. Prior human-centered studies in industry have shown that such urgent issues are treated differently from routine development work \cite{hanna2025behind}. Nevertheless, the code-level characteristics of hot fixing, specifically the technical interactions between developers as reflected in code changes and PR reviews, remain unexplored, particularly in the presence of agentic systems.

\section{Limitations and Future Directions}
\label{limitationsAndFutureWork}


One of the findings in our study is that the distribution of author types is not balanced, with human-authored hot fix PRs outnumbering agent-authored ones. This is somewhat expected given both the dataset composition and the practical reality that high-stakes hot fixes are more likely to be handled by human developers than delegated to autonomous agents. The LLM-based classifiers introduce misclassifications, as they are experimental proxies to critical-scenario triage. We mitigate this threat by applying temporal heuristics based on hot fix characteristics and by manually evaluating 20\% of the classified data with disagreements resolved through discussion.
Finally, our evaluation is limited to the AIDev dataset and hot fix practices may vary across developer communities and repositories.

Future work should explore long-term effects of hot fixes, like defect recurrence and technical debt accumulation, and examine whether agent-authored hot fixes differ in long-term quality. Further research could investigate hybrid workflows, where agents assist humans during emergencies, and evaluate how trust, accountability, and risk perception evolve in such settings.
Overall, our study establishes that critical issues reshape both human and agent behaviour and highlights that autonomous agents already participate 
in urgent software repair. Understanding and guiding this participation is essential as agentic systems become increasingly embedded in production software development workflows.\looseness=-1

\section{Conclusions}
\label{conclusions}

We provide the first large-scale, program source-level examination of hot fix practices and the first comparison of human and autonomous agentic (bots) behaviour in urgent software repair. Using LLM-based classification and temporal analysis, we identify hot fix issues across thousands of software development projects and show that such updates follow structural patterns that differ markedly from regular maintenance. Our findings establish a baseline for understanding hot fixing in the wild and inform the future design and evaluation of autonomous AI agents for production-critical workflows (hot fix bots).
\textbf{Ethical Considerations: }This work has received ethics approval from UCL.



\bibliographystyle{splncs04}
\bibliography{references}

@inproceedings{pinna2026,
title={Comparing {AI} Coding Agents: A Task-Stratified Analysis of Pull Request Acceptance},
author = {Pinna, Giovanni and Gong, Jingzhi and William, David and Sarro, Federica},
booktitle = "MSR 2026"
}

@inproceedings{ssbse:hotcat:2025,
title = "{HotCat}: Green and Effective Feature Selection for HotFix Bug Taxonomy",
author = "{Luis de la Cal} and others",
comment_author = "{Luis de la Cal} and Yazhuo Cao and Irmak Ercevik and Giovanni Pinna and Luke Twist and David Williams and Karine Even-Mendoza and Langdon, William B. and Menendez, Hector D. and Federica Sarro",
year = "2025",
comment_month = nov,
day = "2",
language = "English",
booktitle = "SSBSE",
comment_booktitle = "The 17th Symposium on Search-Based Software Engineering 2025 (SSBSE 2025)",
}

@inproceedings{ASE:refuzzer:2025,
title = "{ReFuzzer}: Feedback-Driven Approach to Enhance Validity of {LLM}-Generated Test Programs",
author = "Iti Shree and Karine Even-Mendoza and Tomasz Radzik",
booktitle = "ASE 2025",
comment_booktitle = "The 40th IEEE/ACM International Conference on Automated Software Engineering, ASE 2025",
comment_year = "2025",
comment_month = nov,
day = "10",
language = "English",
}

@inproceedings{hanna2025behind,
  title={Behind the Hot Fix: Demystifying Hot Fixing Industrial Practices at {Z{\"u}hlke} and Beyond},
  author={Carol Hanna and others},
  comment_author={Hanna, Carol and Elliman, David and Emmerich, Wolgang and Sarro, Federica and Petke, Justyna},
  booktitle={FSE 2025},
  comment_booktitle={Proceedings of the 33rd ACM International Conference on the Foundations of Software Engineering},
  comment_year={2025},
  pages={411--421},
}

@misc(li2025aidev,
      author={{Hao Li} and others},
  comment_author={{Hao Li} and {Haoxiang Zhang} and Ahmed E. Hassan},
  title =       {The Rise of {AI} Teammates in Software Engineering {(SE)} 3.0: How Autonomous Coding Agents Are Reshaping Software Engineering},
  howpublished = {arXiv:2507.15003},
comment_year={2025},
  comment_month =       {20 July},
  note =        {},
      archivePrefix={arXiv},
      primaryClass={cs.SE},
  comment_url={https://arxiv.org/abs/2507.15003}, 
  size =        {22 pages},
  notes =       {Source documentation for LZH's AIDev MSR 2026 challenge track
  
  Also known as \cite{li2025riseaiteammatessoftware}


Queen's University, Kingston, ON, Canada},
)

@article{hanna2024lirev,
author = {Carol Hanna and others},
comment_author = {Hanna, Carol and Clark, David and Sarro, Federica and Petke, Justyna},
title = {Hot Fixing Software: {A} Comprehensive Review of Terminology, Techniques, and Applications},
journal = {TOSEM},
comment_journal = {ACM Trans. Softw. Eng. Methodol.},
year = {2025},
comment_month = dec,
comment_publisher = {ACM},
issn = {1049-331X},
doi = {10.1145/3786330},
}

@misc{hanna2025hotbugsjarbenchmarkhotfixes,
      title={{HotBugs.jar}: {A} Benchmark of Hot Fixes for Time-Critical Bugs}, 
      author={Carol Hanna and Federica Sarro and Mark Harman and Justyna Petke},
      year={2025},
      eprint={2510.07529},
      archivePrefix={arXiv},
      primaryClass={cs.SE},
      url={https://arxiv.org/abs/2510.07529}, 
}

@INPROCEEDINGS{10298540,
  author={Hanna, Carol and Petke, Justyna},
  booktitle={ASE 2023}, 
  comment_booktitle={2023 38th IEEE/ACM International Conference on Automated Software Engineering (ASE)}, 
  title={Hot Patching Hot Fixes: Reflection and Perspectives}, 
  comment_year={2023},
  volume={},
  number={},
  pages={1781--1786},
  doi={10.1109/ASE56229.2023.00021}}

@article{10.1145/3633453,
author = {Albert Ziegler and others},
comment_author = {Ziegler, Albert and Kalliamvakou, Eirini and Li, X. Alice and Rice, Andrew and Rifkin, Devon and Simister, Shawn and Sittampalam, Ganesh and Aftandilian, Edward},
title = {Measuring {GitHub} {Copilot's} Impact on Productivity},
journal = {Commun. ACM},
year = {2024},
issue_date = {March 2024},
publisher = {ACM},
address = {NY, USA},
volume = {67},
number = {3},
month = feb,
pages = {54--63},
numpages = {10},
issn = {0001-0782},
doi = {10.1145/3633453},
}

@INPROCEEDINGS{10992485,
  author={{Zhi Chen} and {Lingxiao Jiang}},
  booktitle={SANER 2025}, 
  title={Evaluating Software Development Agents: Patch Patterns, Code Quality, and Issue Complexity in Real-World {GitHub} Scenarios}, 
  comment_year={2025},
  volume={},
  number={},
  pages={657--668},
  doi={10.1109/SANER64311.2025.00068}}

@article(watanabe2025useagenticcodingempirical,
      title={On the Use of Agentic Coding: An Empirical Study of Pull Requests on {GitHub}}, 
      author={Miku Watanabe and others},
  comment_author={Miku Watanabe and {Hao Li} and Yutaro Kashiwa and Brittany Reid and Hajimu Iida and Ahmed E. Hassan},
  journal =     {TOSEM},
  year =        {},
  volume =      {},
  number =      {},
  pages =       {},
  month =       {},
  note =        {accepted},
  ISSN =        {1049-331X},
  doi =         {10.1145/3798166},
  size =        {},
  notes =       {},
  arxive_year={2026},
  version = {v3},
      eprint={2509.14745},
      archivePrefix={arXiv},
      primaryClass={cs.SE},
  comment_url={https://arxiv.org/abs/2509.14745}, 
)

@misc{Jiang2025,
      title={Agentic Software Issue Resolution with Large Language Models: A Survey}, 
      author={Zhonghao Jiang and David Lo and Zhongxin Liu},
      year={2025},
      eprint={2512.22256},
      archivePrefix={arXiv},
      primaryClass={cs.SE},
      url={https://arxiv.org/abs/2512.22256}, 
comment = {from jiang2025agenticsoftwareissueresolution},
}

@misc{roychoudhury2025agenticaisoftwarethoughts,
      title={Agentic {AI} for Software: thoughts from Software Engineering community}, 
      author={Abhik Roychoudhury},
      year={2025},
      eprint={2508.17343},
      archivePrefix={arXiv},
      primaryClass={cs.SE},
      url={https://arxiv.org/abs/2508.17343}, 
}

@misc{alam2026aiagentinvolvedpull,
      title={Why Are {AI} Agent Involved Pull Requests (Fix-Related) Remain Unmerged? {An} Empirical Study}, 
      author={Khairul Alam and Saikat Mondal and Banani Roy},
      year={2026},
      eprint={2602.00164},
      archivePrefix={arXiv},
      primaryClass={cs.SE},
      url={https://arxiv.org/abs/2602.00164}, 
}

@misc{ehsani2026aicodingagentsfail,
      title={Where Do {AI} Coding Agents Fail? {An} Empirical Study of Failed Agentic Pull Requests in {GitHub}}, 
      author={Ramtin Ehsani and others},
  comment_author={Ramtin Ehsani and Sakshi Pathak and Shriya Rawal and Abdullah Al Mujahid and Mia Mohammad Imran and Preetha Chatterjee},
      year={2026},
      eprint={2601.15195},
      archivePrefix={arXiv},
      primaryClass={cs.SE},
      url={https://arxiv.org/abs/2601.15195}, 
}

@inproceedings{10.1145/3542929.3563482,
author = {{Supriyo Ghosh} and others},
  comment_author = {Ghosh, Supriyo and Shetty, Manish and Bansal, Chetan and Nath, Suman},
title = {How to Fight Production Incidents? {An} Empirical Study on a Large-scale Cloud Service},
comment_year = {2022},
isbn = {9781450394147},
comment_publisher = {ACM},
booktitle = {SoCC 2022},
comment_booktitle = {Proceedings of the 13th Symposium on Cloud Computing},
pages = {126--141},
numpages = {16},
comment_address = {San Francisco, USA},
comment_series = {SoCC '22},
comment_doi = {10.1145/3542929.3563482},
}

@article{10.1093/cybsec/tyab023,
    author = {Roumani, Yaman},
    title = {Patching zero-day vulnerabilities: an empirical analysis},
    journal = {Journal of Cybersecurity},
    volume = {7},
    number = {1},
    pages = {tyab023},
    year = {2021},
    month = {11},
    issn = {2057-2085},
    doi = {10.1093/cybsec/tyab023},
}

@INPROCEEDINGS{HPC:NotHotFix:10174003,
  author={Kalam Azad, Md Abul and others},
  comment_author={Kalam Azad, Md Abul and Iqbal, Nafees and Hassan, Foyzul and Roy, Probir},
  booktitle={MSR 2023}, 
  title={An Empirical Study of High Performance Computing {(HPC)} Performance Bugs}, 
  comment_year={2023},
  volume={},
  number={},
  pages={194--206},
  comment_doi={10.1109/MSR59073.2023.00037},
}

@inproceedings{10.1145/3338906.3338916,
  author = {Domenico Cotroneo and others},
  comment_author = {Cotroneo, Domenico and De Simone, Luigi and Liguori, Pietro and Natella, Roberto and Bidokhti, Nematollah},
title = {How Bad Can a Bug Get? {An} Empirical Analysis of Software Failures in the {OpenStack} Cloud Computing Platform},
booktitle = {FSE 2019},
comment_year = {2019},
isbn = {9781450355728},
comment_publisher = {ACM},
pages = {200--211},
numpages = {12},
comment_address = {Tallinn, Estonia},
comment_series = {ESEC/FSE 2019},
comment_doi = {10.1145/3338906.3338916},
}

@ARTICLE{DoSinOS:5010241,
  author={Gligor, Virgil D.},
  journal={IEEE TSE}, 
  title={A Note on Denial-of-Service in Operating Systems}, 
  year={1984},
  volume={SE-10},
  number={3},
  pages={320--324},
  doi={10.1109/TSE.1984.5010241}}

@article{10.1007/s10207-025-01164-3,
author = {Alqahtani, Norah and Almukaynizi, Mohammed},
title = {{VulnScore}: A deployed system for patch prioritization combining human input and temporal threat intelligence},
year = {2025},
issue_date = {Jan 2026},
publisher = {Springer-Verlag},
address = {Berlin, Heidelberg},
volume = {25},
number = {1},
issn = {1615-5262},
doi = {10.1007/s10207-025-01164-3},
journal = {Int. J. Inf. Secur.},
month = nov,
numpages = {12},
}

@INPROCEEDINGS{9970537,
  author={Costa, Thiago Figueiredo and Tymburiba, Mateus},
  booktitle={SIN 2022}, 
  comment_booktitle={2022 15th International Conference on Security of Information and Networks (SIN)}, 
  title={Challenges on prioritizing software patching}, 
  comment_year={2022},
  volume={},
  number={},
  size = {8 pages},
  address = {Sousse, Tunisia},
  doi={10.1109/SIN56466.2022.9970537}}

@inproceedings{10.1145/3524459.3527343,
author = {Kang, Sungmin and Yoo, Shin},
title = {Language models can prioritize patches for practical program patching},
comment_year = {2022},
isbn = {9781450392853},
comment_publisher = {ACM},
booktitle = {APR 2022},
comment_booktitle = {Proceedings of the Third International Workshop on Automated Program Repair},
pages = {8--15},
numpages = {8},
comment_address = {Pittsburgh, USA},
comment_series = {APR '22},
doi = {10.1145/3524459.3527343},
}

\end{document}